\documentclass[conference]{IEEEtran}
\usepackage{array} 
\usepackage{amsmath}
\usepackage{algorithm}
\usepackage{algpseudocode}
\usepackage{subcaption}
\usepackage{makecell}
\usepackage{multirow}
\DeclareCaptionType{custombox}[\textbf{Prompt}][]
\DeclareCaptionType{custombox1}[\textbf{Example}][]
\usepackage[framemethod=TikZ]{mdframed}
\usepackage{url} 
\usepackage{booktabs}
 \usepackage{graphicx} 
\usepackage{pgfplots}
\pgfplotsset{compat=1.17}
\usepackage{subcaption}
\usepackage{graphicx}
\usepackage{parskip}
\usepackage{authblk}
\usepackage{amsmath}
\AtBeginDocument{%
  }
\usepackage{array}
\usepackage{colortbl}
\usepackage{xcolor}

\newcommand{\applycolormap}[1]{%
  \ifdim #1 pt > 50pt\relax
    \cellcolor{red!#1}\fi#1%
}

\newcommand{\applycolormapmulti}[1]{%
  \ifdim #1 pt > 50pt\relax
    \cellcolor{cyan!#1}\fi#1%
}

\ifCLASSINFOpdf
\else
\fi
\begin{document}
%
\title{MoRSE: Bridging the Gap in Cybersecurity Expertise with Retrieval Augmented Generation}

\author[1]{Marco Simoni}
\author[2]{Andrea Saracino}
\author[3,4]{Vinod Puthuvath}
\author[3,4]{Maurco Conti}

\affil[1]{Istituto di Informatica e Telematica, Consiglio Nazionale Delle Ricerche, Pisa, Italy}
\affil[2]{TeCIP, Scuola Universitaria Superiore Sant’Anna, Pisa, Italy}
\affil[3]{University of Padua, Italy and Delft University of Technology, Netherlands}
\maketitle

\begin{abstract}
In this paper, we introduce MoRSE (Mixture of RAGs Security Experts), the first specialised AI chatbot for cybersecurity. MoRSE aims to provide comprehensive and complete knowledge about cybersecurity.
MoRSE uses two RAG (Retrieval Augmented Generation) systems designed to retrieve and organize information from multidimensional cybersecurity contexts. MoRSE differs from traditional RAGs by using parallel retrievers that work together to retrieve semantically related information in different formats and structures. 
Unlike traditional Large Language Models (LLMs) that rely on Parametric Knowledge Bases, MoRSE retrieves relevant documents from Non-Parametric Knowledge Bases in response to user queries. Subsequently, MoRSE uses this information to generate accurate answers. In addition, MoRSE benefits from real-time updates to its knowledge bases, enabling continuous knowledge enrichment without retraining. We have evaluated the effectiveness of MoRSE against other state-of-the-art LLMs, evaluating the system on 600 cybersecurity specific questions.  
The experimental evaluation has shown that the improvement in terms of relevance and correctness of the answer is more than 10\% compared to known solutions such as GPT-4 and Mixtral 7x8.
\end{abstract}



%

\section{Introduction}

The increasing frequency and sophistication of new cyber threats have made cybersecurity a critical priority across all sectors, with a 15\% increase in three years~\footnote{\url{https://bit.ly/3zxkf2y}} on data breaches only. In recent years, the amount of cybersecurity-related information has exploded, providing important resources to protect against these threats by mitigating risk and improving cybersecurity measures. However, this rapid proliferation of information has led to a cluttered and often unstructured data landscape, complicating the task of deriving actionable insights for professionals~\cite{shin2020review, kokulu2019matched}. In fact, a timely, accurate, and comprehensive understanding of vulnerabilities, exploits, and defense tactics is crucial, as the promptness of such information can significantly impact cybersecurity decisions~\cite{sun2023cyber, kaur2023artificial}. Recently, large language models (LLMs) have become important tools for synthesizing huge amounts of information in various fields, including cybersecurity~\cite{sarker2024llm}. However, their reliability varies for technical topics where inaccuracies are critical~\cite{kandpal2023large, yao2024survey, kocon2023chatgpt, ray2023chatgpt}. LLMs can produce \textit{hallucinatory} responses, meaning that they produce answers that are not true or reliable, especially struggling with the dynamic and evolving nature of cyber threats~\cite{ji2023survey, kasai2024realtime, mallen-etal-2023-trust}. This problem is particularly pronounced in code generation tasks, where LLMs often produce non-functioning code for complex queries~\cite{liu2024your}. Specifically, when the model does not know the correct answer, hallucinations are inevitable (\textit{Epistemic}~\cite{yadkori2024believebelievellm}). This can happen if the model does not have enough training data or if its capacity is limited. An example of hallucination in cybersecurity is shown in Example~\ref{example_ground_truth} (ground truth) and Example~\ref{example_gpt4} (the answer of GPT-4) to the question: \textit{What is CVE-2017-5162?}. This shows that GPT-4 is not able to provide the correct answer. To overcome these limitations, it is crucial to build continuous learning mechanisms into LLMs that allow them to dynamically update their knowledge base with the latest information~\cite{shuster-etal-2021-retrieval-augmentation}. Although retraining can be time-consuming, updating with new information is essential to maintain the relevance of LLM responses~\cite{izacard2020leveraging, pmlr-v162-borgeaud22a}. For this reason, companies are quickly discovering the value of \textit{Retrieval Augmented Generation} chatbots. 
Although RAG systems have made significant advances on cybersecurity, researchers and practitioners often lacks comprehensive models that unify diverse cybersecurity data and dynamic, real-time updated systems that are essential for addressing the evolving cyber threat landscape. In fact, effective analysis of threats, vulnerabilities and exploits requires a thorough understanding of individual entities and their broader interactions.

\captionof{custombox1}{\textbf{Ground Truth.}}
\label{example_ground_truth}
\begin{mdframed}[roundcorner=10pt, leftmargin=1, rightmargin=1, linecolor=black, outerlinewidth=1]
\textbf{CVE:} CVE-2017-5162\\
\textbf{Description:} An issue was discovered in BINOM3 Universal Multifunctional Electric Power Quality Meter. Lack of authentication for remote service gives access to application set up and configuration.
\end{mdframed}


\captionof{custombox1}{\textbf{GPT-4 0125-Preview Response.}}
\label{example_gpt4}
\begin{mdframed}[roundcorner=10pt, leftmargin=1, rightmargin=1, linecolor=black, outerlinewidth=1]
\textbf{CVE:} CVE-2017-5162\\
\textbf{Description:} CVE-2017-5162 is a vulnerability associated with Broadcom Wi-Fi chipsets. It is one of the vulnerabilities part of the BroadPwn exploit. BroadPwn is a critical security flaw that allows a remote attacker to compromise a vulnerable device over the air, without any interaction from the user.
\end{mdframed}

\paragraph{Contribution} In this paper, we present MoRSE (Mixture of RAGs Security Experts), a novel framework developed for Question and Answering tasks in the field of cybersecurity. Unlike other chatbots 
MoRSE is the first RAG framework that aims to provide comprehensive coverage of cybersecurity knowledge.
To achieve this, MoRSE uses two RAG systems in cascade.
These two RAGs divide the process into two distinct phases: \textit{Information Retrieval} phase, which is managed by \textit{Multiple Retrievers}, followed by the answer generation phase, which is driven by \textit{Large Language Models}. The first RAG, called \textit{Structured RAG}, comprises retrievers that are tailored to fast retrieval tasks, as they retrieve information from preprocessed, structured data. The second, \textit{Unstructured RAG}, is slower due to its higher complexity, but retrieves a larger amount of information as it enables the exploration of data in its original form. The second RAG is only activated if the first RAG does not find any relevant documents for a specific query. Each retriever is specialized in a specific area of cybersecurity and collects data from key resources such as
{MITRE}\footnote{https://attack.mitre.org/}, {CVE repositories}\footnote{https://cve.mitre.org/},
{Metasploit}\footnote{https://www.metasploit.com/}, and {ExploitDB}\footnote{https://www.exploit-db.com/}. The name of our system, MoRSE, is in fact inspired by the \textit{Mixture of Experts} (MoE) paradigm~\cite{masoudnia2014mixture}, reflecting the specialized skills of the individual retrievers.
In addition, MoRSE receives new knowledge through real-time updates of its retrievers' knowledge bases, which enables a continuous expansion of knowledge. Indeed, a key advantage of non-parametric memory models such as RAG is the ability to update knowledge at test time. In contrast, Parametric-only models require the entire model to be retrained, which can be both time-consuming and resource-intensive.

 Using a comprehensive set of 600 cybersecurity questions, consisting of 150 General Cybersecurity questions, 150 Multi-Hop (i.e., questions involving multiple entities contained in different documents) Cybersecurity questions, and 300 CVE questions (on vulnerabilities), we evaluated MoRSE alongside other commercial, well-known LLMs, including {GPT-4}\footnote{https://chat.openai.com/}, {GEMINI}\footnote{https://gemini.google.com/app}, MIXTRAL~\cite{jiang2024mixtral}, and {HACKERGPT}\footnote{https://chat.hackerai.co/}. The questions were classified based on the \textit{Diamond Model}~\cite{Caltagirone2013TheDM}, which we used to create questions that are representative of the real needs of the cybersecurity world. Two experts validated the ground truth of the questions, with a Cohen's Kappa~\cite{Kappa1, Kappa2} index as a measure of agreement between the two experts equal to $0.82$ out of 1.  

The comparative analysis shows the superior performance of MoRSE for cybersecurity queries. In terms of relevance and correctness of answers~\cite{es-etal-2024-ragas}, MoRSE outperforms other models by more than \(15\%\) for General questions and by more than \(10\%\) for Multi-Hop questions and CVE Questions. In terms of accuracy, MoRSE outperforms GPT-4 by 50\% for CVE Questions. This confirms its effectiveness in specialized domains. We validated these results with the \textit{LLM as a Judge} method~\cite{zheng2024judging}, which derives Elo ratings for each model and confirms MoRSE's leading performance compared to all competitors.

The main contributions of our work can be summarized as follows:
\begin{itemize}
   \item We introduce MoRSE, an open-source framework\footnote{\url{https://github.com/Mixture-of-RAGs-Security-Experts/MoRSE}} which is the first attempt to integrate two RAG systems to handle multidimensional cybersecurity contexts. This architecture enables a unique synthesis of different data sources and improves the depth and relevance of security insights.
    
   \item We introduce a three-part evaluation test suite that measures the relevance, similarity and correctness of RAG systems in conjunction with LLMs. We have further validated these results with two additional test suites based on the \textit{LLM as a Judge} approach. To the best of our knowledge, we are the first to provide such contribution.

    
  \item We demonstrate how MoRSE can exploit its unique real-time cybersecurity keyword detection capability to improve the correctness of responses by 10\% compared to GPT-4, addressing the critical need for timely and accurate security analysis.

  \item MoRSE differs from conventional RAGs by using parallel retrievers that work together to retrieve semantically related information in different formats and structures. This is particularly important in the cybersecurity domain, where different data types such as exploit code, TTP descriptions, CVEs and white papers often exist for a particular threat but are rarely related to each other. MoRSE exploits these parallel retrievers and LLMs to integrate related information and provide comprehensive answers to queries.
    
\end{itemize}

\paragraph{Organisation} The rest of the article is structured as follows. Section~\ref{sec:bg} discusses background information on LLMs and RAGs. Section~\ref{sec:me} describes the MoRSE architecture. Section~\ref{sec:ex} describes the experiments performed to evaluate the performance of MoRSE, including a comparative analysis with known commercial models. Section~\ref{sec:rw} provides an overview of related research work that includes various cybersecurity tools such as knowledge graphs, entity extraction tools, chatbots, and cyber threat intelligence (CTI). The conclusions and future work directions are presented in Section~\ref{sec:cf}.

\section{background}\label{sec:bg}
This section provides an overview of the basic concepts necessary for understanding the architecture of MoRSE.
\subsection{Large Language Models}

Large language models (LLMs) represent a significant advancement in the field of Natural Language Processing (NLP) and are based on the Transformer model~\cite{vaswani2017attention}. These models are trained on large text datasets and are able to generate coherent and contextually relevant texts based on input prompts. The capabilities of LLMs go beyond text generation and include tasks such as language translation, summarizing, answering questions, etc.

The introduction of models such as GPT~\cite{radford2018improving} and BERT~\cite{devlin2018bert} has demonstrated the potential of LLMs to revolutionize language understanding and generation through unsupervised and bidirectional training~\cite{radford2018improving, devlin2019bert}. With the development of GPT-3~\cite{brown2020language}, the scalability of these models reached new heights, illustrating their ability to perform a wide range of NLP tasks without task-specific training.

Despite their advantages, LLMs face several challenges. Ethical considerations, such as the spread of bias and misinformation, are a major concern~\cite{bender2021dangers}. In addition, the environmental impact of training and operating these computationally intensive models has raised questions about their sustainability~\cite{strubell2019energy}. Efforts to overcome these challenges include research into more efficient training methods and models that are able to understand and generate texts with greater accuracy and less bias~\cite{wang2020efficient}.

\subsection{Retrieval Augmented Generation}
Retrieval-Augmented Generation~(RAG) combines traditional language models with external databases to improve natural language processing (NLP) tasks~\cite{lewis2020retrieval, pmlr-v119-guu20a, karpukhin2020dense}. RAG models use a retriever to retrieve relevant information and a generator to generate answers based on the retrieved information. This improves accuracy and relevance, especially for domain-specific queries~\cite{borgeaud2021improving, izacard2020leveraging}.

RAG's strengths lie in its ability to update knowledge bases without retraining and customise components for specific tasks such as cybersecurity~\cite{lewis2020retrieval, pmlr-v119-guu20a}.

However, RAG struggles with latency and scalability issues, especially when processing concurrent queries~\cite{lewis2020retrieval}. Despite these limitations, RAG remains a versatile tool for a range of NLP applications, from chatbots to content creation. Ongoing research focuses on optimizing retrieval mechanisms and computational power~\cite{borgeaud2021improving, izacard2020leveraging}.
\subsection{Definitions}
We report in the following a set of definitions which will be used in the rest of the paper:

\begin{itemize}
     \item \textbf{Retriever} is a component that identifies and retrieves relevant information or documents from a knowledge base. This process is essential to provide the necessary context and content that LLM uses to generate accurate and informed answers~\cite{karpukhin2020dense, zhou2015learning}.
    
     \item \textbf{Knowledge Base} is a repository of information that the retriever accesses to find relevant data or documents. This is fundamental to the system's ability to retrieve contextually relevant content, essential for generating well-informed and accurate answers~\cite{li2020cskb}.
    
     \item \textbf{Embeddings} are numerical representations of text that assign a low dimension to a term. Within this context, embedding vectors of analogous terms exhibit proximity, encapsulating semantic meaning. This facilitates the comparison between queries and the knowledge base~\cite{muennighoff2022mteb}.

 \item \textbf{Context} refers to the relevant information or data retrieved by the system, which surrounds and informs a particular query. The model requires this contextual information to formulate answers that are precise, comprehensive, and directly linked to the content found in the knowledge base~\cite{ram2023context}.
    
     \item \textbf{Prompt} refers to the structured input that is created from the retrieved context, which is then fed into the generative model. This prompt guides the model in generating a coherent, contextually relevant response that directly addresses the user's request~\cite{liu2023pre}.
    
     \item \textbf{Semantic Similarity} evaluates how closely the content of a user's query matches the information in the knowledge base, focusing on meaning rather than word-for-word matching. This evaluation guarantees the relevance and accuracy of the retrieved data and supports the generative model in creating appropriate answers. In MoRSE, we use \textit{Cosine Similarity}~\cite{chandrasekaran2021evolution} to measure the proximity of embedding vectors because it has a high correlation with human judgment~\cite{semscore}.
    \item \textbf{Multi-hop queries} are defined as requests for information that necessitate indirect reasoning over multiple pieces of interconnected data. They typically arise in complex question-answering tasks where a single piece of evidence is insufficient to resolve the query, and the system must \textit{hop} across different data points or documents to piece together a response. 
\end{itemize}

\section{MoRSE Architecture}\label{sec:me}

This section describes the structure of the MoRSE system in detail. It explains the function of the individual components and how they interact to process a request and generate a response. We used the Langchain framework \footnote{https://www.langchain.com/} to develop the MoRSE architecture. The Table~\ref{table:symbol_value_type} contains the key to the symbols used in this explanation to facilitate the understanding of the following discussion. The first two symbols, ($\alpha$) and ($\beta$), are embedding models, while the last two, ($\gamma$) and ($\theta$), are Transformer models.
\begin{table}[h]
\centering
\caption{Mapping of Symbols to Values.}
\label{table:symbol_value_type}
\small 
\setlength{\tabcolsep}{5pt} 
\renewcommand{\arraystretch}{1.2} 
\begin{tabular}{|c|c|}
\hline
\textbf{Symbol} & \textbf{Value}                      \\ \hline
$\alpha$       & \texttt{BAAI/bge-large-en}~\cite{zhang2023retrieve} \\ 
$\beta$        & \texttt{BAAI/bge-large-en-v1.5}~\cite{zhang2024soaring,li2023making,zhang2023retrieve}\\
$\gamma$       & \texttt{valhalla/t5-base-e2e-qg}~\cite{raffel2020exploring} \\ 
$\theta$       & \texttt{dslim/bert-large-NER}~\cite{devlin2018bert} \\ \hline

\end{tabular}
\end{table}

\subsection{MoRSE Overview}
\begin{figure}[H]
  \centering
  \includegraphics[scale=0.1]{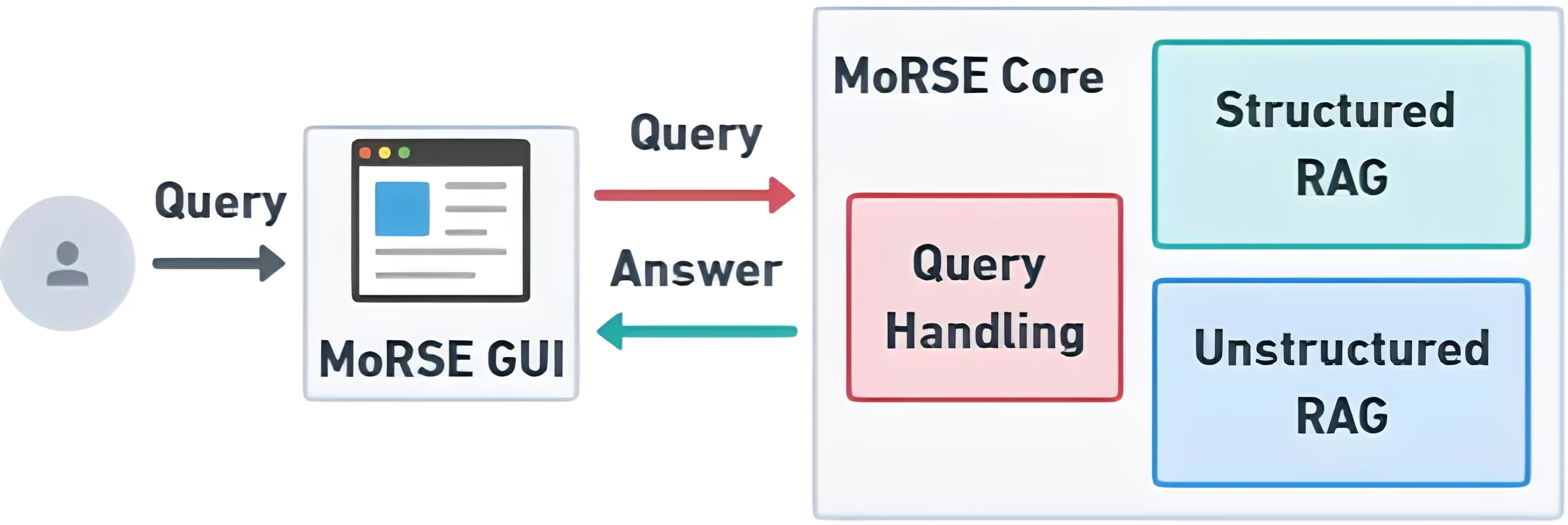}
  \caption{MoRSE Overview.}
  \label{fig:GUI}
\end{figure}

As shown in Figure~\ref{fig:GUI}, \textbf{MoRSE} consists of two main components: a graphical user interface (GUI) and the MoRSE core. The GUI enables interaction with the user by allowing the input of queries and displaying the answers in a structured way\footnote{\url{https://github.com/Mixture-of-RAGs-Security-Experts/MoRSE}}.

The MoRSE core consists of three key components, which in turn manage the user query and compose the answer:

\begin{itemize}
 \item \textbf{Query Handling Module:} This module performs the pre-processing of user queries and specializes in the management of multi-hop queries and complex questions, especially in the context of Common Vulnerability and Exposures (CVEs) and Common Weakness Enumerations (CWEs).
 
 \item \textbf{Structured RAG:} The first of the two RAGs is composed by retrievers that retrieve information from pre-processed, structured data. The pre-processing phase involves converting chunks of text from various sources that are part of the knowledge base, such as academic papers and cybersecurity websites, into well-defined structures. These structures are designed to contain generated questions and contextualized entity descriptions that facilitate the precise retrieval of information in response to user queries.
 \item \textbf{Unstructured RAG:} This RAG is used if the structured RAG could not find a suitable answer. It searches for information in unstructured and unprocessed raw text that belongs to its knowledge base. Accessing unstructured data allows the exploration of data in its original form without the limitations imposed by preprocessing, thus providing a wider range of search options in exchange for a higher response time. This type enables the exploration of data in its original form without the restrictions imposed by preprocessing.
\end{itemize}

The RAGs will compose the answer to each query and return it to the GUI for structured visualization. In the following we will detail the components of the MoRSE Core.

\subsection{MoRSE Core}

\begin{figure*}[!ht]
  \centering
  \includegraphics[scale=0.5]{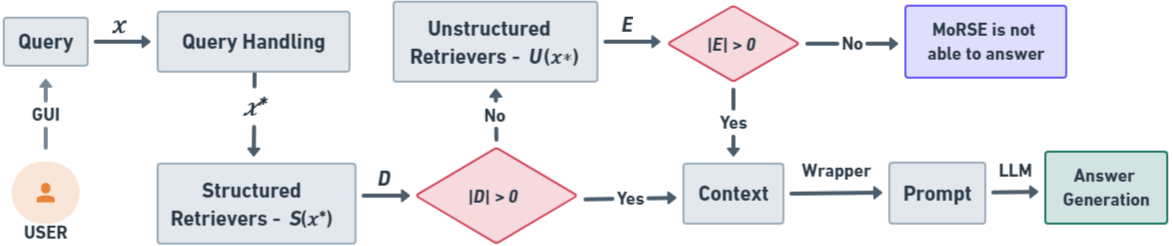}
  \caption{MoRSE Core Workflow.}
  \label{fig:core}
\end{figure*}
\paragraph{\textbf{MoRSE Core Workflow}} Figure~\ref{fig:core} shows the first stage of MoRSE Core process, starting with the \textit{Query Handling} model. This module converts the original query \(x\) into an optimized version \(x^*\) (see subsection~\ref{sec:query_handling}). First, \(x^*\) is forwarded to the \textit{Structured RAG} module for processing. The structured RAG path, denoted as $\mathcal{S}$, begins with the \textit{Structured Retrievers}, focused on high accuracy and fast responses to efficiently process most queries. Their primary function, $\mathcal{S}(x^*)$, is to identify and retrieve information pertinent to the query.

When activated, the structured retrieval process, which is executed via \(\mathcal{S}(x^*)\), assigns a set of potentially relevant documents from a predefined \textit{Knowledge Base} to the query \(x^*\). In particular, \(\mathcal{D} = \text{top-k}(\mathcal{S}(x^*))\) represents the selection of the top \(k\) documents that \(\mathcal{S}\) considers most relevant for the query based on a similarity score. If \(\mathcal{D}\) is not empty (\(|\mathcal{D}| > 0\)), this means that a relevant context has been found. The workflow then proceeds to use this context and moves on to the next phase, where the retrieved information (\(\mathcal{D}\)) is wrapped in a \textit{Prompt}, which is used by \textit{LLM} to generate a response.

If the \textit{Structured Retrievers} do not yield relevant documents (\(|\mathcal{D}| = 0\)), the workflow moves to the unstructured path and calls the \textit{Unstructured Retrievers}, denoted as \(\mathcal{U}\). At this stage, \(\mathcal{E} = \text{top-k}(\mathcal{U}(x^*))\) represents the set of documents retrieved by \(\mathcal{U}\), which are designed to process complex queries that are not readily covered by structured data patterns.
After a successful retrieval of relevant information in one of the two ways — indicated by \(|\mathcal{D}| > 0\) for structured retrieval or \(|\mathcal{E}| > 0\) for unstructured retrieval —the \textit{Wrapper} module integrates the acquired context and generates a prompt for the Large Language Model (LLM). The LLM then performs \textit{Answer Generation}, creating a detailed response to the user's question.

\begin{figure}[!ht]
  \centering
  \includegraphics[scale=0.5]{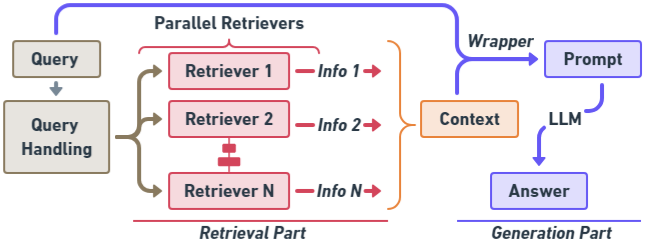}
  \caption{RAG Architecture Overview.} 
  \label{fig:rag_structure}
\end{figure}

\paragraph{\textbf{RAG Architecture}} The RAG architecture of the MoRSE system, which is used in both the Structured~(\ref{sec:structured}) and Unstructured RAG~(\ref{sec:unstructured}), follows the same underlying logic shown in Figure~\ref{fig:rag_structure}. This architecture is divided into two parts: 
\begin{enumerate}
    \item The \textbf{Retrieval} part, which consists of \textit{Parallel Retrievers} used to collect relevant information for the query.

    \item  The \textbf{Generation} part, in which the Large Language Model (LLM) uses the context provided in the \textit{Prompt} to generate responses. After the retrieval phase, the collected information (info $1$ to info $N$) is merged into a \textit{Context}, which is \textit{Wrapped}, along with the user query, in a \textit{Prompt} used by LLM to generate the \textit{Answer}. The logic of the architecture is formalized in the Algorithm~\ref{alg:rag_multi_retriever}.
\end{enumerate}

\begin{algorithm}
\caption{RAG with $N$ Parallel Retrievers}\label{alg:rag_multi_retriever}
\begin{algorithmic}[1]
\Require User query $Q$
\Ensure Answer $A$
\State \textbf{Predefined:} Retrievers $r_1, r_2, \ldots, r_N$
\Procedure{ExecuteRAG}{$Q$}
    \State Initialize context set $C \gets \emptyset$
    \For{$i \gets 1$ \textbf{to} $N$} \Comment{In parallel for each $i$-th retriever}
        \State $I_i \gets r_i(Q)$ \Comment{Retrieve information using $i$-th retriever}
        \State Sort $I_i$ by relevance scores to find the most pertinent documents
        \State $I_{\text{top}} \gets \text{top}_k(I_i)$ \Comment{Select the top-k documents to form a new set $I_{\text{top}}$}
        \If{not empty($I_{\text{top}}$)}
            \State $C \gets C \cup \{I_{\text{top}}\}$ \Comment{Incorporate the top-k documents into the context $C$}
        \EndIf
    \EndFor
    \If{$C = \emptyset$}
        \State \Return \textit{"No relevant information found."}
    \EndIf
    \State $P \gets \text{wrap}(C, Q)$ \Comment{Construct a prompt from the aggregated context C and the User query Q.}
    \State $A \gets LLM(P)$ \Comment{Use the Large Language Model to generate an answer.}
    \State \Return $A$
\EndProcedure
\end{algorithmic}
\end{algorithm}

\subsection{Query Handling}
\label{sec:query_handling}
This component improves the intelligence of the MoRSE system by managing complex query types and enriching the context. Below are the specific functions and the composition of this component:

\subsubsection{Functionalities}
\begin{itemize}
    \item \textbf{Multi-Hop Question Handling}: Deals with queries involving multiple related entities and allows the system to handle and answer complex multi-hop questions. Existing Retrieval Augmented Generation systems struggle with multi-hop queries due to their design limitations and the lack of a dedicated benchmark dataset for this type of query~\cite{tang2024multihoprag, xiong2020answering}.
 \item \textbf{Context Enriching}: Generates additional questions from each identified entity, expanding and enriching the context available for generating informed answers.
 \item \textbf{Solving the CVE-CWE Conundrum}: Effectively handles queries related to Common Vulnerabilities and Exposures (CVE) and Common Weakness Enumerations (CWE), which are challenging for generative models due to their technical complexity~\cite{abdallah2023generator, gao-etal-2023-precise}.
\end{itemize}

\subsubsection{Components}
\begin{itemize}
  \item \textbf{User Query}: Initiates the process when a user submits a query via the graphical user interface.
 \item \textbf{CVE-CWE Keyword Extraction}: Extracts keywords related to CVEs and CWEs when a query is received. 
 \item \textbf{Get CVE Description}: Retrieves detailed descriptions of CVEs, including information about vulnerabilities, affected software and finders.
 \item \textbf{Get CWE Description}: Retrieves descriptions of CWEs that provide information about the type of software vulnerabilities, potential impact and mitigation strategies.
 \item \textbf{Entity Extractor}: Utilizes the Haystack framework~\footnote{https://haystack.deepset.ai/} with the \texttt{$\theta$} model to identify and extract relevant entities (people or concepts) from user queries, improving the system’s ability to handle Multi-Hop queries.
\end{itemize}

The complete workflow is outlined in Algorithm~\ref{alg:query_handling}, demonstrating the mechanism of Query Handling within the MoRSE system.

\begin{algorithm}
\caption{Query Handling Process}\label{alg:query_handling}
\algnewcommand\algorithmicinput{\textbf{Input:}}
\algnewcommand\INPUT{\item[\algorithmicinput]}
\algnewcommand\algorithmicoutput{\textbf{Output:}}
\algnewcommand\OUTPUT{\item[\algorithmicoutput]}
\begin{algorithmic}[1]  

\INPUT A user query $q$ on a specific cybersecurity topic.
\OUTPUT A series of refined queries $Q'$ for in-depth analysis.

\Procedure{VulnerabilityExtractor}{$q$}
    \State Extract keywords $K$ related to CVE and CWE from $q$.
    \State Retrieve detailed descriptions $D$ for each keyword in $K$ from relevant databases.
    \State Update $q$ by substituting each keyword in $K$ with its corresponding description from $D$, yielding $q'$.
    \State \Return $q'$
\EndProcedure

\Procedure{EntityExtractor}{$q'$}
    \State Utilize the Haystack framework with "dslim/bert-base-NER" to detect entities $E$ in $q'$.
    \State Start with an empty \texttt{Queries\_List} $Q$ and include $q'$.
    \For{each detected entity $e \in E$} 
        \If{$e$ refers to a person (PER)}
            \State Append a question about $e$ to $Q$: \texttt{"Who is $e$?"}
        \ElsIf{$e$ pertains to an object or concept (OBJ/CON)}
            \State Append a question about $e$ to $Q$: \texttt{"What is $e$?"}
        \EndIf
    \EndFor
    \State \Return $Q$
\EndProcedure

\Procedure{QueryHandling}{$q$}
     \State $q_{vuln} \gets \Call{VulnerabilityExtractor}{q}$
    \State $Q' \gets \Call{EntityExtractor}{q_{vuln}}$
    \State \textbf{return} $Q'$
\EndProcedure
\end{algorithmic}
\end{algorithm}

\subsection{Structured RAG}
\label{sec:structured}
\begin{figure}[!ht]
  \centering
  \includegraphics[scale=0.245]{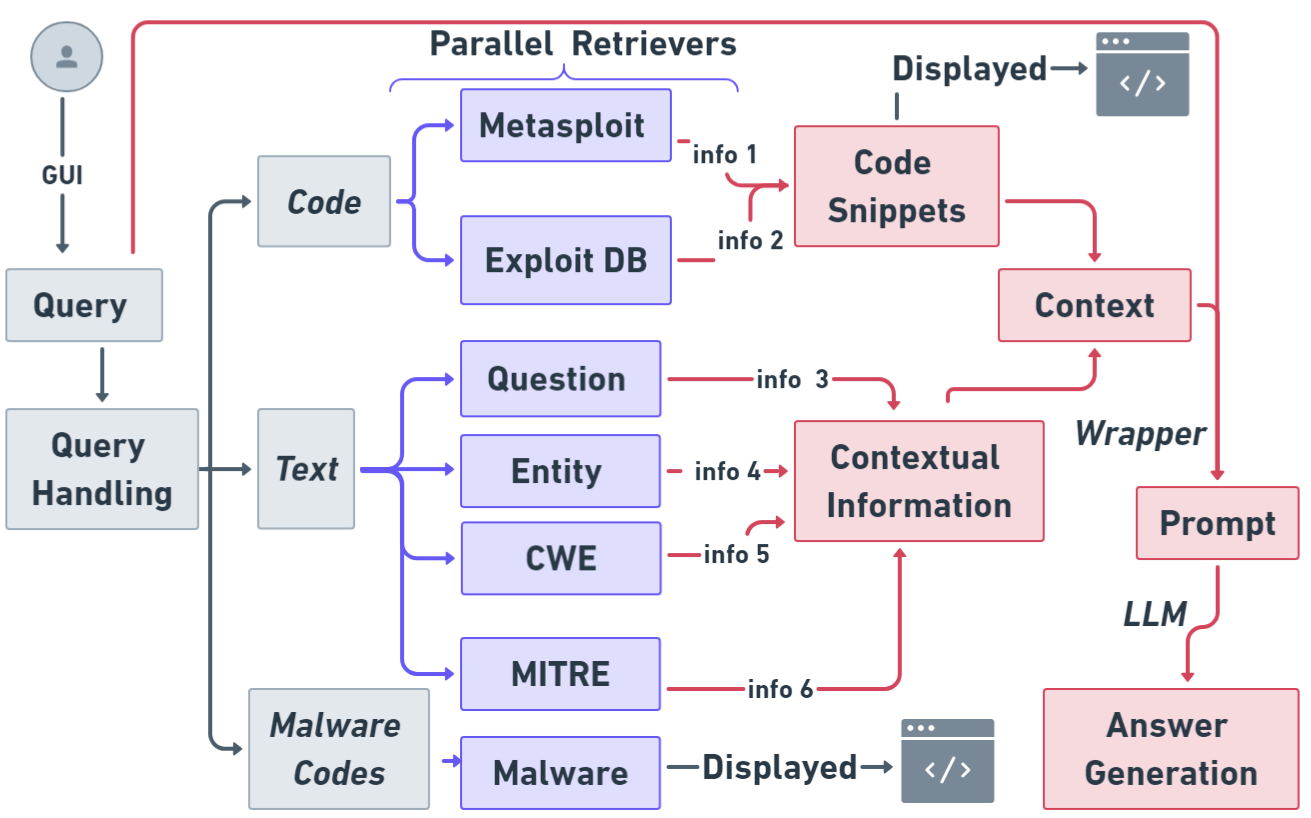}  
  \caption{Structured RAG Workflow.}
  \label{fig:firstlayer}
\end{figure}

As illustrated in Figure~\ref{fig:firstlayer}, the Structured RAG module works post-\textit{Query Handling} by forwarding refined queries to seven \textit{Parallel Retrievers}, called \textit{Structured Retrievers}, each of which specializes in specific cybersecurity topics. Given a query, the information contained in the knowledge base of a Retriever is inserted into the \textit{Context} if its similarity to the query is above a predefined threshold. In order to establish the threshold for each retriever, we conducted an analysis on the scores of top 50 results from a series of test queries. Thresholds were then determined by assessing the distribution of scores \footnote{https://github.com/Mixture-of-RAGs-Security-Experts/MoRSE/tree/main/Retriever-Threshold-Scores}. 
In particular, we used the median value of the test distributions as threshold for the \textit{MITRE Retriever} and the \textit{Malware Retriever}, as these typically retrieve shorter texts. For \textit{Question Retrieval System}, \textit{CWE Retriever}, \textit{Metasploit Retriever} and \textit{Entity Retriever}, we chose the third quartile (Q3) of the test distributions as threshold, as they generally retrieve longer texts. The \textit{ExploitDB Retriever} works without a threshold and uses the TF-IDF algorithm~\cite{ramos2003using}.
To mitigate embedding biases~\cite{brunet2019understanding}, we used two different embeddings for the retrievers, ($\alpha$) and ($\beta$). The following paragraphs delineate each retriever's functionalities.
\paragraph{\textbf{Mitre Retriever}}
The knowledge base of this retriever comes from the website of the MITRE Corporation\footnote{https://attack.mitre.org/software/}. It is structured as a graph database containing two primary node categories: \textit{Malware} and \textit{Techniques}. Each \textit{Malware} node in the database contains a name and a description of MITRE.
We create \textit{Technique} nodes, which consist of technique names and descriptions, by collecting and analysing technique-related links from the MITRE website. This retriever utilizes embeddings (\texttt{$\alpha$}). To ensure accurate matches, the system exclusively evaluates malware with a similarity score exceeding 0.7, corresponding to the Test distribution's median.

\paragraph{\textbf{Metasploit Retriever}}
We have developed the Metasploit Retriever so that it can be effectively integrated into the Metasploit Framework. Its knowledge base includes over 4900 cybersecurity elements, including exploits, encoders, payloads, and various modules. To increase the retrieval speed, we only index salient parts of codes such as code descriptions and exploit information. It performs a semantic search using a similarity value of 0.75 (Q3 of the test distribution) with (\texttt{$\alpha$}) embedding, along with a keyword search supported by the TF-IDF algorithm~\cite{ramos2003using}.


\paragraph{\textbf{ExploitDB Retriever}}
The knowledge base of this retriever is made up of exploits from the ExploitDB framework. These codes often lack descriptions, so we use a keyword search with the TF-IDF algorithm~\cite{ramos2003using} to focus only on important data such as CVE identifiers and author names, which are usually at the beginning of the scripts. To increase the retrieval speed, we only index the first 600 characters of each script, as this information is often contained in the first sections of the code.

\paragraph{\textbf{Question Retrieval System}}

This system acts as a knowledge base containing questions extracted from chunks of the original documents to better select the most important parts of the document and the explanations contained therein. A user query is compared with these questions and if there is a match, the chunk from which the matching question was extracted is retrieved. During preprocessing, documents are divided into 2000-character chunks. Model (\texttt{$\gamma$}) generates about seven questions per chunk, which are refined with \textit{Mistral-7B-Instruct-v0.2}~\cite{jiang2023mistral} for better alignment. The system uses (\texttt{$\beta$}) embeddings and deploys four retrievers in sequence, each selecting the ten most relevant documents. The results are merged, filtered with a similarity threshold of 0.6 (Q3 of the test distribution) and reordered based on the \textit{Lost In the Middle} Principle~\cite{liu2024lost}, by placing key information at the beginning or end of the context. Redundant questions targeting the same document chunk are removed to streamline the context.


\paragraph{\textbf{Entity Retriever}} 
This retriever contains entities from document chunks together with their descriptions extracted from the context of the chunk. During the pre-processing phase, we segment the documents into 500-word chunks. This segmentation enables the model (\texttt{$\theta$}) to identify and classify relevant entities more precisely. 
The contextual descriptions for each entity are then created using the \textit{mistralai/Mistral-7B-Instruct-v0.2} model and converted by (\texttt{$\beta$}) embeddings into a searchable format that is retrievable with a similarity threshold of 0.5 (Q3 of the test distribution).



\paragraph{\textbf{Malware Retriever}}
This retriever contains more than 1000 malware source codes originating from GitHub pages. The Malware Retriever uses a semantic search with (\texttt{$\alpha$}) embeddings and a threshold value of 0.7 (median of the test distribution) to match search queries with malware names. In case of matches, all related files are displayed on the graphical interface. 

\paragraph{\textbf{CWE Retriever}} The CWE Retriever uses a semantic search with (\texttt{$\alpha$}) embeddings to match user queries with CWE descriptions, as described in Section~\ref{sec:query_handling}. Operating with a threshold of 0.7 (Q3 of the test distribution), it showcases the 10 most relevant CWEs. When a query closely aligns with a CWE, the retriever presents detailed information, including code examples.

\paragraph{\textbf{Context Construction}} When creating the final context for the prompt, inputs from two main sources (code snippets and contextual information) are organized to ensure visibility and impact during the \textit{Generation Phase}:
\begin{itemize}
    \item \textbf{Code Snippets}: Following the \textit{Lost In the Middle principle}~\cite{liu2024lost}, code snippets from \textit{Metasploit} and \textit{ExploitDB} are prioritized at the beginning of the prompt so that they are immediately visible.

 \item \textbf{Contextual Information}: To ensure that essential and concise information is presented to the LLM, the content from the queries \textit{Mitre}, \textit{CWE} and \textit{Entity} retrievers is placed at the end of the prompt. The more comprehensive outputs of the Question Retriever, which is equipped with a reordering function that respects the \textit{Lost In the Middle principle}~\cite{liu2024lost}, are placed in the middle.
\end{itemize}


\subsection{Unstructured RAG}
\label{sec:unstructured}
\begin{figure}[!ht]

  \centering
  \includegraphics[scale=0.18]
  {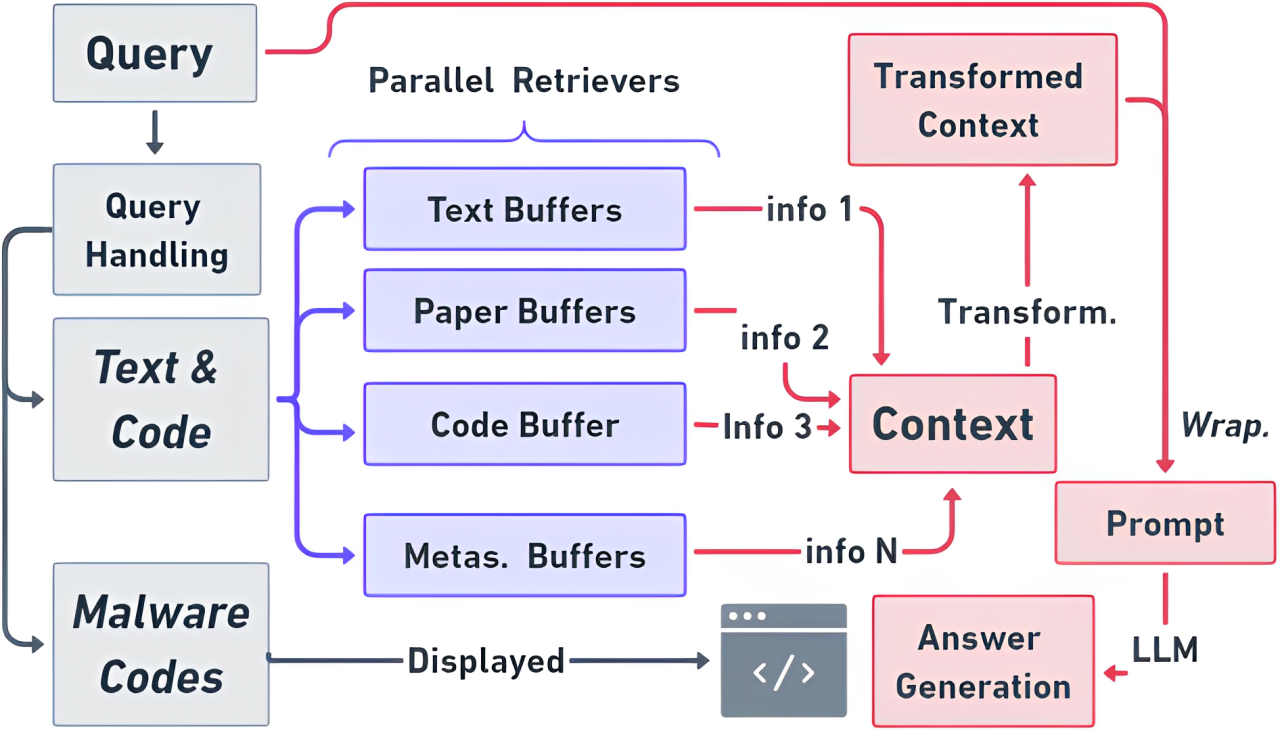}  
  \caption{Unstructured RAG Workflow}
  \label{fig:Structured RAG}
\end{figure}

The unstructured RAG, shown in Figure \ref{fig:Structured RAG}, plays a crucial role in the MoRSE system by handling cybersecurity queries that the structured RAG cannot solve. The module utilizes retrievers, called \textit{buffers}, to store documents in chunks of 2000 characters while maintaining the integrity of the original information. All buffers work as hybrid retrievers that use both semantic search and keyword search with the BM25 algorithm~\cite{wang2021bert}. Unlike other configurations, these retrievers do not have a fixed threshold for semantic search; instead, they are configured to return the top five documents regardless of similarity scores. This decision enables further \textit{Context Transformation} process to apply semantic thresholds, ensuring the flexibility and comprehensiveness of the retrieval process.

\paragraph{\textbf{Classification of Buffers}}
Buffers are categorized according to the type of data they process, and there are four different types of buffers:

\begin{itemize}
 \item \textbf{Text Buffers}: Process content from websites and blogs with \textbf{five} separate buffers, each analyzing data using the (\textit{\texttt{$\alpha$}}) embedding.
 \item \textbf{Metasploit Buffers}: \textbf{five} buffers containing entire codes from the Metasploit framework that uses the (\textit{\texttt{$\alpha$}}) embedding for effective processing.
 \item \textbf{Code Buffers}: A single buffer processes code snippets from Exploit DB and also uses (\textit{\texttt{$\alpha$}}) embedding for optimal analysis.
 \item \textbf{Paper Buffers}: Academic papers are managed by \textbf{three} buffers that use the (\textit{\texttt{$\beta$}}) embedding to better handle the complex language typical of academic content~\cite{bge_embedding}. This choice is based on the higher performance values of embedding, which indicate better retrieval capabilities. 
\end{itemize}

The process \textbf{Context Transformation} refines information from buffers through four  phases:
\begin{itemize}
    \item \textbf{Splitting Stage:} Document are split into 300-character chunks, improving relevance selection and reducing noise from larger chunks.
    \item \textbf{Redundant Content Removal:} In this phase, \texttt{$\beta$} embeddings are used to remove redundant content from the segmented chunks and improve the clarity and uniqueness of the output.
    \item  \textbf{Filtration Stage:} Relevant data chunks are selected using (\texttt{$\beta$}) embeddings with a threshold of 0.6 set to the third quartile (Q3) of similarity results from tests with 156 queries to ensure relevance from the knowledge base.
    \item  \textbf{Reordering Phase:} Finally, the data is ordered according to the (\textit{Lost in the Middle}) principle to prioritize the visibility of important information in the response.
\end{itemize}

\begin{table*}[h!]
\centering
\caption{Comparative Assessment of RAG Models on 156 General and 150 Multi-Hop Cybersecurity Questions}
\label{table:combined_performance}
\begin{tabular}{lcccccc|cccccc}
\toprule
& \multicolumn{6}{c}{\textbf{General Cybersecurity Questions}} & \multicolumn{6}{c}{\textbf{Multi-Hop Cybersecurity Questions}} \\
\cmidrule(lr){2-7} \cmidrule(lr){8-13}
\textbf{Model} & \multicolumn{2}{c}{\textbf{Relevance}} & \multicolumn{2}{c}{\textbf{Similarity}} & \multicolumn{2}{c}{\textbf{Correctness}} & \multicolumn{2}{c}{\textbf{Relevance}} & \multicolumn{2}{c}{\textbf{Similarity}} & \multicolumn{2}{c}{\textbf{Correctness}} \\
& \(\mu\) & \(\sigma\) & \(\mu\) & \(\sigma\) & \(\mu\) & \(\sigma\) & \(\mu\) & \(\sigma\) & \(\mu\) & \(\sigma\) & \(\mu\) & \(\sigma\) \\
\midrule
\textbf{MoRSE}          & \textbf{0.90} & \textbf{0.05} & \textbf{0.95} & \textbf{0.02} & \textbf{0.71} & \textbf{0.08} & \textbf{0.93} & \textbf{0.05} & \textbf{0.93} & \textbf{0.03} & \textbf{0.70} & \textbf{0.09} \\
\textbf{GPT-4 0125-Preview} & 0.74 & 0.35 & 0.93 & 0.03 & 0.59 & 0.07 & 0.75 & 0.33 & 0.92 & 0.03 & 0.62 & 0.10 \\
\textbf{MIXTRAL 7X8}        & 0.77 & 0.43 & 0.92 & 0.03 & 0.58 & 0.08 & 0.60 & 0.42 & 0.92 & 0.04 & 0.61 & 0.09 \\
\textbf{HACKERGPT}          & 0.66 & 0.33 & 0.92 & 0.04 & 0.58 & 0.08 & 0.73 & 0.18 & 0.79 & 0.05 & 0.47 & 0.03 \\
\textbf{GEMINI 1.0 Pro}     & 0.61 & 0.43 & 0.91 & 0.05 & 0.58 & 0.08 & 0.70 & 0.37 & 0.90 & 0.07 & 0.58 & 0.10 \\
\bottomrule
\end{tabular}
\end{table*}


\section{EXPERIMENTS and EVALUATION}\label{sec:ex}

The evaluation of Retrieval Augmented Generation systems and Large Language Models in the field of cybersecurity is particularly challenging due to their dual role in information retrieval and content generation. The lack of standardized benchmarks covering a wide range of real-world operational cyber tasks complicates the evaluation of LLMs for cybersecurity~\cite{es-etal-2024-ragas, gennari2024considerations, sultana2023towards}.

Key evaluation challenges include verifying the accuracy of the retrieved information, the effectiveness of its use by LLM, and the overall quality of the content generated. Traditional methods that focus on language comprehension may not adequately reflect real-world performance~\cite{wang-etal-2023-knn}.

To effectively address these challenges, we developed a three-part evaluation strategy for MoRSE and compared its performance with other known LLMs and RAG systems in answering cybersecurity questions. MoRSE was compared to competing models such as GPT-4 0125-Preview, MIXTRAL, HACKERGPT, and GEMINI 1.0 Pro. The three different evaluation test suites are:
\begin{itemize}
    \item Using the RAGAS framework~\cite{es-etal-2024-ragas}, we evaluate MoRSE's responses against a ground truth using a set of metrics.
     \item Using a method proposed by Zheng et al.~\cite{zheng2024judging}, we computed Elo Ratings for MoRSE and competing models by reference-guided pairwise comparison using \textit{GPT-4 0125-Preview} as a judge. This provides a quantitative measure of relative performance.
     \item Following Zheng et al.~\cite{zheng2024judging}, \textit{GPT-4 0125-Preview} also rates the responses on a scale of 1 to 5 based on their comparison with the ground truth references, which have 5 as the highest default score.
\end{itemize}
We conducted the assessment using three different types of cybersecurity questions. The first category, \textit{General Cybersecurity Questions}, includes 150 simple one-liners that address a wide range of cybersecurity topics. The second category, \textit{Multi-Hop Cybersecurity Questions}, includes 150 complex queries that require a thorough, multi-layered understanding. The third category focuses on 300 \textit{Common Vulnerability Exposure} (CVE) questions that address specific security vulnerabilities. The questions were classified based on the \textit{Diamond Model}~\cite{Caltagirone2013TheDM}, which we used to create questions representative of real cybersecurity world needs, and the sample size was statistically selected based on standard methods~\footnote{\url{https://github.com/Mixture-of-RAGs-Security-Experts/MoRSE/tree/main/CohensKappaEvaluation}}. Two experts with 12 and 2 years of experience validated the ground truth. We used Cohen's Kappa~\cite{Kappa1, Kappa2} as a metric to evaluate the agreement between the two experts. They categorized the answers as \texttt{[Correct]}, \texttt{[Incorrect]} or \texttt{[Partially\_Correct]} depending on the context of the questions. The experts agreed well (Cohen's Kappa = 0.82), indicating an \textit{Almost Perfect} agreement~\cite{cohens}.

For all three evaluation test suites, we generated the answers from MoRSE using the \textit{mistralai/Mistral-7B-Instruct-v0.2} model, operating on an NVIDIA A100 80GB GPU.

\subsection{First Test Suite: Ground Truth Assessing alignment using the RAGAS framework}
Using the RAGAS framework~\cite{es-etal-2024-ragas}, we focused on three metrics: \textit{answer relevance}, \textit{answer similarity}, and \textit{answer correctness}. To calculate these metrics, we used \textit{GPT-4 0125-Preview} as the underlying model for all calculations.

\textbf{Answer Relevance}, shown in Equation~\ref{eq:relevancy}, measures how pertinent the generated answer is to the given prompt. It is calculated by generating related questions from the model's answer and comparing their embeddings to the original question using cosine similarity:
\begin{equation}
    \text{Answer Relevance} = \frac{1}{N} \sum_{i=1}^{N} \cos(\mathbf{E}_g^i, \mathbf{E}_o),
\label{eq:relevancy}
\end{equation}
where \(\mathbf{E}_g^i\) and \(\mathbf{E}_o\) are the \(\beta\)-embeddings of the generated and original questions, respectively, and \(N\) equal to 3, is the number of generated questions. 

\textbf{Answer Similarity}, shown in Equation~\ref{eq:answer_similarity}, evaluates semantic congruence between model-generated responses and predefined correct answers, calculated as:
\begin{equation}
\text{Answer Similarity} = \frac{\mathbf{V}_{\text{ground truth}} \cdot \mathbf{V}_{\text{generated}}}{\|\mathbf{V}_{\text{ground truth}}\| \|\mathbf{V}_{\text{generated}}\|},
\label{eq:answer_similarity}
\end{equation}
where $\mathbf{V}_{\text{ground truth}}$ and $\mathbf{V}_{\text{generated}}$ represent vector representations of the ground truth and generated answers, respectively.

\textbf{Answer Correctness}, shown in Equation~\ref{eq:answer_correctness} and ~\ref{eq:answer_correctness_2}, evaluates the factual accuracy of generated answers against ground truth. It combines semantic similarities and factual correctness:
\begin{equation}
AC = w_{FC} \cdot FC + w_{SS} \cdot SS,
\label{eq:answer_correctness}
\end{equation}
where \(FC\) is the factual correctness, quantified using the F1 score that considers True Positives (\(TP\)), False Positives (\(FP\)), and False Negatives (\(FN\)):
\begin{equation}
FC = \frac{|TP|}{|TP| + 0.5 \cdot (|FP| + |FN|)},
\label{eq:answer_correctness_2}
\end{equation}
and \(SS\) is the semantic similarity between the generated and ground truth answers. \(w_{FC}\) and \(w_{SS}\) are the weights assigned to \(FC\) and \(SS\), respectively 0.75 and 0.25. 
In order to calculate TP, FP and FN, RAGAS framework uses the following prompt instruction: \textit{Extract the following from the given question and ground truth: ``TP'': statements that are present in both the answer and the ground truth, ``FP'': statements present in the answer but not found in the ground truth, ``FN'': relevant statements found in the ground truth but omitted in the answer}.

Each of these three metrics requires an embedding model to compute distances between sentences and a large language model (LLM) for evaluating answer relevance and correctness. We chose \textit{GPT-4 0125-Preview} as the LLM and (\textit{\texttt{$\beta$}}) as the embedding model.



\label{sec:cybersecurity_questions}
\paragraph{\textbf{Performance Analysis on General and Multi-Hop Questions}}
Table~\ref{table:combined_performance} shows the results of MoRSE and the other models for \textit{General Cybersecurity Questions} and \textit{Multi-Hop Cybersecurity Questions}. The metrics for each model are expressed as mean (\(\mu\)) and standard deviation (\(\sigma\)), which indicate the average performance and variability, respectively.

\par \textbf{Insights from General Cybersecurity Questions:}
For the general cybersecurity questions, MoRSE showed superior performance in all metrics, with a mean relevance score of 0.90, a similarity score of 0.95, and a correctness score of 0.71, indicating a high degree of agreement between the answers and the query prompts, as well as factual accuracy. In comparison, all other models showed lower consistency and effectiveness, especially in terms of correctness.

\par \textbf{Insights Multi-Hop Cybersecurity Questions:}
When evaluating complex multi-hop cybersecurity queries, the MoRSE model outperforms its competitors and proves that it is capable of answering complicated questions. The data shows that MoRSE scores consistently high on all metrics, with average scores of 0.93 for relevance and similarity and 0.70 for correctness.
Other models show significant performance degradation, particularly in correctness, with GPT-4 0125-Preview achieving a mean score of 0.62 and MIXTRAL 0.61, indicating a lower capacity to handle multi-hop questions.

\paragraph{\textbf{Performance Analysis on CVE Questions}}
Table~\ref{tab:cve_queries} shows how MoRSE and GPT-4 0125 preview approach 300 CVE queries. We chose GPT-4 0125 preview because it performed best as the second model in both general and multi-hop contexts (see Table~\ref{table:combined_performance}). We focus on answer similarity and correctness metrics for scoring answers because they are strictly based on ground truth and answer relevance does not measure factuality. Moreover, we compute \textit{Accuracy} metric, calculated by checking whether the models correctly identified the vulnerability in the given queries. Regarding Correctness, the MoRSE model scored 0.64 because its responses often include explanation of related exploit codes, which are not present in the ground truth that only describes the vulnerability specifics. This extra information, while useful, lowers the correctness score as it deviates from the expected response. GPT-4 0125-Preview lags behind in this domain-specific challenge. MoRSE achieved an accuracy of 84\%, surpassing the GPT-4 0125-Preview model, which had an accuracy of 34\%. Our comparison reveals that MoRSE significantly outperforms GPT-4 0125-Preview in accurately identifying vulnerabilities. 


\begin{table}[!ht]
\centering
\caption{Performance comparison of models on 300 CVE Queries.}
\begin{tabular}{@{}lcccccc@{}}
\toprule
\textbf{Model} & \multicolumn{2}{c}{\textbf{Similarity}} & \multicolumn{2}{c}{\textbf{Correctness}} & \textbf{Accuracy} \\ 
\cmidrule(lr){2-3} \cmidrule(lr){4-5}
& \textbf{$\mu$} & \textbf{$\sigma$} & \textbf{$\mu$} & \textbf{$\sigma$} & \\ 
\midrule
\textbf{MoRSE} & 0.903 & 0.028 & 0.640 & 0.111 & 84\% \\
GPT-4 0125-Preview & 0.852 & 0.004 & 0.558 & 0.107 & 34\% \\
\bottomrule
\end{tabular}
\label{tab:cve_queries}
\end{table}
We cannot calculate the accuracy metric for general and multi-hop queries because, unlike CVE queries, they lack strictly factual data points, such as a specific vulnerability to identify. For CVEs, accuracy is straightforward: we check if the model identified the correct vulnerability. In contrast, general and multi-hop questions often lack such clear data and require assessment based on multiple aspects depending on the question type.

\subsection{Retrievers Impact analysis}
\begin{figure}[ht!]
    \centering
    \begin{subfigure}[b]{0.5\textwidth}
        \centering
        \begin{tikzpicture}
            \begin{axis}[
                ybar,
                symbolic x coords={Malware, Metasploit, ExploitDB, CWE, MITRE, Entity, Q. Ret. Sys.},
                xtick=data,
                x tick label style={rotate=45, anchor=east},
                ylabel={Impact (\%)},
                ymin=0,
                ymax=60,
                bar width=0.4cm,
                nodes near coords,
                nodes near coords align={vertical},
                every node near coord/.append style={font=\footnotesize, /pgf/number format/.cd, fixed, precision=1},
                enlargelimits=0.15,
                width=9cm,
                height=4.5cm,
                xlabel style={font=\small},
                ylabel style={font=\small},
                ymajorgrids=true,
                grid style=dashed,
                area legend
            ]
            \addplot coordinates {(Malware,6.1) (Metasploit,8.6) (ExploitDB,7) (CWE,7) (MITRE,9) (Entity,21.7) (Q. Ret. Sys.,56.3)};
            \end{axis}
        \end{tikzpicture}
        \caption{Impact of Retrievers on General Questions}
        \label{fig:general-questions}
    \end{subfigure}
    
    \begin{subfigure}[b]{0.5\textwidth}
        \centering
        \begin{tikzpicture}
            \begin{axis}[
                ybar,
                symbolic x coords={Malware, Metasploit, ExploitDB, CWE, MITRE, Entity, Q. Ret. Sys.},
                xtick=data,
                x tick label style={rotate=45, anchor=east},
                ylabel={Impact (\%)},
                ymin=0,
                ymax=40,
                bar width=0.4cm,
                nodes near coords,
                nodes near coords align={vertical},
                every node near coord/.append style={font=\footnotesize, /pgf/number format/.cd, fixed, precision=1},
                enlargelimits=0.15,
                width=9cm,
                height=4.5cm,
                xlabel style={font=\small},
                ylabel style={font=\small},
                ymajorgrids=true,
                grid style=dashed,
                area legend,
                bar shift=0pt,
                fill=blue,
            ]
            \addplot coordinates {(Malware,6.3) (Metasploit,12) (ExploitDB,6) (CWE,5) (MITRE,7) (Entity,28.3) (Q. Ret. Sys.,35.4)};
            \end{axis}
        \end{tikzpicture}
        \caption{Impact of Retrievers on MultiHop Questions}
        \label{fig:multi-questions}
    \end{subfigure}
    
    \begin{subfigure}[b]{0.5\textwidth}
        \centering
        \begin{tikzpicture}
            \begin{axis}[
                ybar,
                symbolic x coords={Malware, Metasploit, ExploitDB, CWE, MITRE, Entity, Q. Ret. Sys.},
                xtick=data,
                x tick label style={rotate=45, anchor=east},
                ylabel={Impact (\%)},
                ymin=0,
                ymax=35,
                bar width=0.4cm,
                nodes near coords,
                nodes near coords align={vertical},
                every node near coord/.append style={font=\footnotesize, /pgf/number format/.cd, fixed, precision=1},
                enlargelimits=0.15,
                width=9cm,
                height=4.5cm,
                xlabel style={font=\small},
                ylabel style={font=\small},
                ymajorgrids=true,
                grid style=dashed,
                area legend,
                bar shift=0pt,
                fill=green
            ]
            \addplot coordinates {(Malware,1) (Metasploit,31) (ExploitDB,18) (CWE,10) (MITRE,5) (Entity,10) (Q. Ret. Sys.,14)};
            \end{axis}
        \end{tikzpicture}
        \caption{Impact of Retrievers on CVE Questions}
        \label{fig:cve-questions}
    \end{subfigure}
    
    \caption{Impact of Retrievers on General, MultiHop and CVE Questions}
\end{figure}
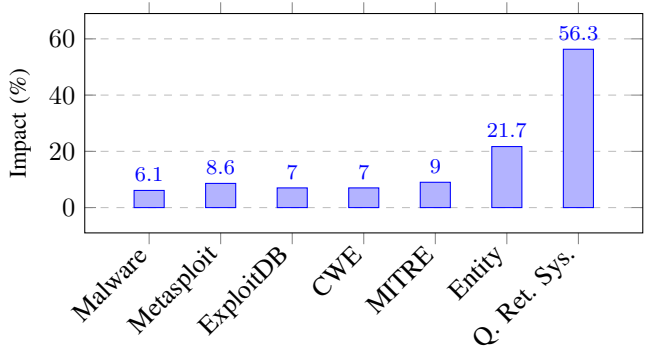
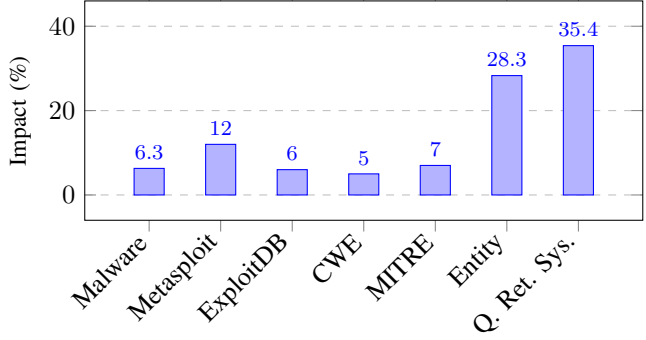
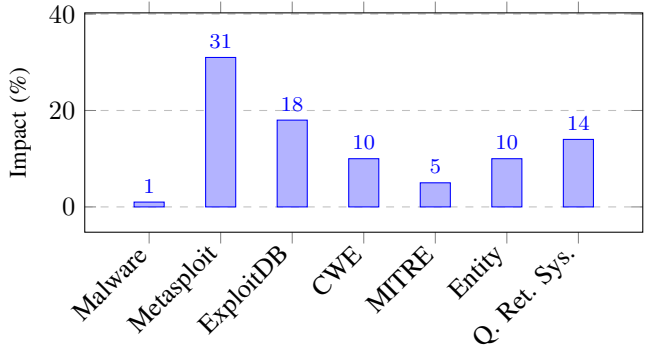

To calculate the impact of each retriever on 600 questions, we applied a systematic methodology. First, we collected all contexts generated for 150 general questions, 150 multi-hop questions and 300 CVE questions. We then analyzed the frequency with which each retriever was able to successfully retrieve relevant information within these contexts. The frequency of successful retrievals for each retriever was then calculated as a percentage of the total questions in each category. In this way, we were able to quantify the performance and impact of each retriever in both general and multi-hop question scenarios. Figures~\ref{fig:general-questions},~\ref{fig:multi-questions} and  ~\ref{fig:cve-questions} show the impact of the different retrievers for each question category. For general questions, the Question Retrieval System has the highest impact at 56.3\%, followed by the Entity Retriever with 21.7\%. Other retrievers, such as MITRE, CWE, ExploitDB, Metasploit and Malware Retriever, have an impact of between 6.1\% and 9.0\%. For multi-hop questions, the Question Retrieval System significantly influences results with a 35.4\% contribution, and the Entity Retriever also plays an important role with a 28.3\% contribution. The Metasploit Retriever has an increased influence of 12\%. The other retrievers (Malware, CWE, ExploitDB and MITRE) have an influence of between 5\% and 7\%. Figure~\ref{fig:cve-questions} shows the impact on CVE questions, where the ExploitDB and Metasploit Retrievers have the highest impact at 18\% and 31\% respectively. The other retrievers (Malware, CWE, MITRE, Entity, and Question Ret. Sys.) have an impact ranging from 1\% to 14\%.

\subsection{Second Test Suite with LLM as Judge: Reference-Guided Pairwise Comparison}
\captionof{custombox}{\textbf{The prompt for reference-guided pairwise comparison.}}
\label{fig:prompt_1}
\begin{mdframed}
Please act as an impartial judge and evaluate the quality of the responses provided by two AI assistants to the user question displayed below. Your evaluation should consider correctness and helpfulness. You will be given a reference answer, assistant A’s answer, and assistant B’s answer. Your job is to evaluate which assistant’s answer is better. Begin your evaluation by comparing both assistants’ answers with the reference answer. Identify and correct any mistakes. Avoid any position biases and ensure that the order in which the responses were presented does not influence your decision. Do not allow the length of the responses to influence your evaluation. Do not favor certain names of the assistants. Be as objective as possible. After providing your explanation, output your final verdict by strictly following this format: ``[A]'' if assistant A is better, ``[B]'' if assistant B is better, and ``[C]'' for a tie.

\textbf{[User Question]} \textit{Insert user question here.}

\textbf{[The Start of Reference Answer]} \\
\textit{Insert reference answer here.} \\
\textbf{[The End of Reference Answer]}

\textbf{[The Start of Assistant A's Answer]} \\
\textit{Insert Assistant A's answer here.} \\
\textbf{[The End of Assistant A's Answer]}

\textbf{[The Start of Assistant B's Answer]} \\
\textit{Insert Assistant B's answer here.} \\
\textbf{[The End of Assistant B's Answer]}
\end{mdframed}
In our evaluation, we used GPT-4 0125-Preview as a judge to evaluate and compare the performance of MoRSE, GPT-4 0125-Preview, MIXTRAL 7X8, GEMINI 1.0 Pro and HACKERGPT. This method is based on the research proposed by Zheng et al.~\cite{zheng2024judging}, which shows that GPT-4 has a high agreement with human judgment with an 80\% agreement rate. As shown in Prompt~\ref{fig:prompt_1}, given a query and the corresponding \textit{reference response}, we ask GPT-4 0125-Preview to choose the best response between Model A and Model B.
After GPT-4 0125-Preview completed its evaluation of all possible battles among the competing models, documented in Table \ref{tab:battle_count_fraction}, we derived three different metrics for each model: \textit{Elo Rating}, \textit{Bootstrap-enhanced Elo Ratings} and \textit{Maximum Likelihood Estimation}: 
\begin{itemize}
    \item \textbf{Elo Ratings:} it quantifies the comparative skill levels across entities in competitive scenarios, making it apt for evaluating models based on their head-to-head outcomes. This approach involves initial computation using a linear update algorithm, opting for a conservative $K$-factor to ensure stability in ratings, minimizing bias from recent encounters. Equation \ref{eq:elo_1} shows the Elo rating formula used in our context:
    \begin{equation}
    \label{eq:elo_1}
        R_{new} = R_{old} + K \times (S - E),
    \end{equation}
    
    where $R_{new}$ and $R_{old}$ represent the new and old Elo ratings, respectively. The constant $K$, set equal to 4, controls the volatility of the rating, $S$ denotes the actual match outcome (1 for a win, 0.5 for a tie, and 0 for a loss), and $E$ is the expected outcome, as calculated in Equation \ref{eq:elo_2}:
    \begin{equation}
    \label{eq:elo_2}
        E = \frac{1}{1 + 10^{\frac{R_{new} - R_{old}}{400}}}.
    \end{equation}

    \item \textbf{Maximum Likelihood Estimation:} Utilizing logistic regression for MLE, we further analyzed the pairwise comparisons, deducing each model's probability of outperforming another, thereby enriching our evaluative scope with probabilistic insights.
    The logistic regression model used for MLE can be formalized in Equation \ref{eq:mle}:
    \begin{equation}
    \label{eq:mle}
    \log\left(\frac{p}{1-p}\right) = \beta_0 + \beta_1X_1 + \dots + \beta_pX_p,
    \end{equation}
    where \(p\) is the probability of model $A$ winning over model $B$, \(X_i\) represents the feature vector indicating the participation of models in each match, and \(\beta_i\) are the coefficients learned by the regression, correlating with the Elo scores. The computed coefficients \(\beta_i\) are then scaled and adjusted to derive the final Elo scores for each model.

     \item \textbf{Bootstrap-enhanced Elo Ratings:} To address potential biases related to battle sequencing, we employed a Bootstrap approach, enhancing the robustness of Elo ratings and facilitating confidence interval estimations for a more reliable assessment. The Bootstrap method involves: sampling battles, calculating Elo ratings, forming a rating distribution across rounds, and determining confidence intervals (median, 2.5\%, and 97.5\% percentiles).

\end{itemize}

\paragraph{\textbf{Performances on General Cybersecurity Question}} As shown in Table \ref{tab:elo_ratings}, MoRSE leads with the highest Elo score of 1244 and an MLE Elo of 1252.43, indicating superior performance in general cybersecurity knowledge. It has a strong bootstrap score of 1225.54, indicating consistent results within the 1275-1175 confidence interval. It is followed by GPT-4 with an Elo of 1083 and MLE Elo of 1107.07, also showing reliable performance. GEMINI and MIXTRAL show moderate performance, with Elo ratings of 926 and 885, respectively. HACKERGPT performs the worst with the lowest Elo score.

\paragraph{\textbf{Performances on Multi-Hop Cybersecurity Questions}} As shown in Table \ref{tab:elo_ratings}, in Multi-Hop queries, MoRSE again excels with an Elo of 1280 and an MLE Elo of 1323.67. Its bootstrap score of 1170.62 with a narrow interval indicates a high stability of performance. GPT-4 remains a competitive second with an Elo of 1157. MIXTRAL, despite its lower performance on general questions, performs better on Multi-Hop questions with a better Elo of 921, while GEMINI struggles, as reflected in its Elo of 907. HACKERGPT continues to show less stability in this area as well.

\paragraph{\textbf{Complete Questions Insights}} Considering 156 \textit{General Questions} and 150 \textit{Multi-Hop Questions}, MoRSE stands out with an Elo of 1267 and an MLE Elo of 1294.66, complemented by a solid bootstrap score within a trustworthy interval, proving its robustness. GPT-4's performance is also solid with an Elo of over 1100. The performances of GEMINI and MIXTRAL are very similar, with Elo Ratings between 900 and 950. HACKERGPT remains less consistent overall.

\setlength{\tabcolsep}{4pt} 
\begin{table*}[!ht]
\centering
\caption{Fraction of Wins (No Ties) of Each Combination of Models for \textit{General Cybersecurity Questions} and \textit{Multi-Hop Cybersecurity Questions}.} 
\label{tab:battle_count_fraction}
\begin{tabular}{lcccccccc}
\toprule
\multirow{2}{*}{\textbf{Model}} & \multicolumn{2}{c}{\textbf{GPT-4 0125-Preview}} & \multicolumn{2}{c}{\textbf{GEMINI 1.0 Pro}} & \multicolumn{2}{c}{\textbf{MIXTRAL 7X8}} & \multicolumn{2}{c}{\textbf{HACKERGPT}} \\
 & General Qs & Mul.Hop Qs & General Qs & Mul.Hop Qs & General Qs & Mul.Hop Qs & General Qs & Mul.Hop Qs \\
\midrule
\textbf{MoRSE} & \textbf{70\%} & \textbf{64\%}  & \textbf{84\%}  & \textbf{90\%}  & \textbf{88\%}  & \textbf{86\%} \ & \textbf{91\%}  & \textbf{95\%}  \\
\textbf{GPT-4 0125-Preview} & - & - & 67\%  & 83\%  & 88\% & 77\%  & 91\% & 92\% \\
\textbf{GEMINI 1.0 Pro} & - & - & - & - & 55\%  & 40\%  & 70\%  & 70\%  \\
\textbf{MIXTRAL 7X8} & - & - & - & - & - & - & 70\%  & 78\%  \\
\bottomrule
\end{tabular}
\end{table*}

\begin{table*}[!ht]
\centering
\caption{Elo Ratings, MLE Ratings, and Bootstrap Ratings for each competing Models.}

\begin{tabular}{lccc|ccc|ccc}
\toprule
& \multicolumn{3}{c}{\makecell{\textbf{General Cyber. Questions}}} & \multicolumn{3}{c}{\makecell{\textbf{Multi-Hop Cyber. Questions}}} & \multicolumn{3}{c}{\makecell{\textbf{Complete Cyber. Questions}}} \\
\cmidrule(lr){2-4} \cmidrule(lr){5-7} \cmidrule(lr){8-10}
\textbf{Model} & \textbf{Elo} & \textbf{MLE Elo} & \textbf{Bootstrap} & \textbf{Elo} & \textbf{MLE Elo} & \textbf{Bootstrap} & \textbf{Elo} & \textbf{MLE Elo} & \textbf{Bootstrap} \\
\midrule
\textbf{MoRSE}          & 1244 & 1252.43 & 1225.54 (1275-1175)    & 1280 & 1323.67 & 1170.62 (1178-1163) & 1267 & 1294.66 &  1183.81(1192-1175)\\
\textit{GPT-4 0125}     & 1083 & 1107.07 & 1096.08 (1105-1063) & 1157 & 1201.19 & 1086.08 (1116-1048) & 1184 & 1151.37 & 1041.05(1050-932) \\
\textit{GEMINI}         & 926  & 955.67  &  994.53  (1003-984) & 907  & 860.86 & 948.16  (981-917) & 918 & 935.27 &  986.81(996-978) \\
\textit{MIXTRAL}        & 885  & 882.53  & 891.73  (900-884)   & 921  & 941.46  & 929.75   (930-927) & 905 & 907.12 &  945.65(954-935) \\
\textit{HACKERGPT}      & 863  & 802.30  & 875.52  (925-850) & 734  & 672.34  &  846.32 (852-838) & 727 & 711.70 & 841.59(849-833) \\
\bottomrule
\end{tabular}
\label{tab:elo_ratings}
\end{table*}

\subsection{Third Test Suite: LLM as Judge with Five-Level Judgment Criteria Based on Top-Scoring References}
\captionof{custombox}{\textbf{The prompt for Five-Level Judgment Criteria Referencing the Top-Scoring Reference.}}
\label{fig:prompt_2}
\begin{mdframed}
\textit{Task Description}:
A Question, a response to evaluate, a reference answer that gets a score of 5, and a score rubric representing a evaluation criteria are given.
1. Write a detailed feedback that assess the quality of the response strictly based on the given score rubric, not evaluating in general.
2. After writing a feedback, write a score that is an integer between 1 and 5. You should refer to the score rubric.
3. The output format should look as follows: "Feedback: \{\{write a feedback for criteria\}\} [RESULT] \{\{an integer number between 1 and 5\}\}"
4. Please do not generate any other opening, closing, and explanations. Be sure to include [RESULT] in your output.

\textbf{Score Rubrics:}
\textit{Is the response correct, accurate, and factual based on the reference answer?}
\begin{itemize}
    \item \textit{Score 1}: The response is completely incorrect, inaccurate, and/or not factual. 
    \item \textit{Score 2}: The response is mostly incorrect, inaccurate, and/or not factual.
    \item \textit{Score 3}: The response is somewhat correct, accurate, and/or factual.
    \item \textit{Score 4}: The response is mostly correct, accurate, and factual.
    \item \textit{Score 5}: The response is completely correct, accurate, and factual.
\end{itemize}
\textbf{The Question}: \textit{[Question]}\\
\textbf{Response to evaluate}: \textit{[Response to Evaluate]}\\
\textbf{Reference Answer (Score 5)}: \textit{[Reference Answer]}
\end{mdframed}
 
As illustrated in Prompt~\ref{fig:prompt_2}, our methodology involves utilizing GPT-4 0125-Preview to assign scores ranging from 1 to 5 to each model answers based on a Top-Scoring Reference.

These scores are based on a comparison with a reference answer, which is given a perfect score of 5. The procedure for this scoring corresponds to the methodology described in~\footnote{\url{https://huggingface.co/learn/cookbook/rag_evaluation}}.

As can be seen from Table~\ref{tab:5_scores}, MoRSE outperforms in all question categories (General questions, Multi-Hop questions, and CVE questions). GPT-4 0125-Preview follows MoRSE as the second most competent model and shows strong versatility in all categories, but with a significantly lower performance than MoRSE.
Other models, GEMINI, HACKERGPT and MIXTRAL, show different levels of performance, with none reaching the effectiveness of MoRSE or GPT-4 0125-Preview in terms of general cybersecurity knowledge or the more targeted questions on Multi-Hop and CVE.

\begin{table}[!ht]
\caption{Scores Across All Cybersecurity Questions types.}
\centering
\begin{tabular}{@{}lcccc@{}}
\toprule
\textbf{Model} & \textbf{General Qs} & \textbf{Multi-Hop Qs} & \textbf{CVE Qs} \\
\midrule
\textbf{MoRSE} & \textbf{4.54} & \textbf{4.65} & \textbf{4.024} \\
GPT-4 0125 & 3.14 & 3.54 & 2.193 \\
GEMINI & 2.91 & 2.43 & - \\
HACKERGPT & 2.65 & 3.05 & - \\
MIXTRAL & 2.58 & 1.40 & - \\
\bottomrule
\end{tabular}
\label{tab:5_scores}
\end{table}
\subsection{Test Case Analysis for MoRSE RAG Retrievers}

This section reports the evaluation on the performance of both the structured and unstructured RAG components within the MoRSE system, as shown in Table~\ref{tab:perfomances}. We evaluated each retriever against a customized set of 100 test questions based on their specialized knowledge in cybersecurity, focusing on processing time efficiency, dense retriever size, denoted by \textit{Size} and number of documents (\textit{No. Doc}). We also test the reliability of the structured Retrievers using test queries with error rate analysis to estimate future reliability using a confidence interval.

\paragraph{\textbf{Analysis of the performance of Structured Retrievers}}
Table~\ref{tab:perfomances} shows the performance metrics for Structured RAG Retrievers within the MoRSE system. The \textit{ExploitDB Retriever} \textit{Size} is not listed as it uses TF-IDF for retrieval. The \textit{Malware Retriever} processes queries with average times of 0.061 seconds on GPU and 0.110 seconds on CPU. The \textit{CWE Retriever} and the \textit{MITRE Retriever} have average processing times of 0.083 and 0.057 seconds on GPU and 0.108 and 0.13 seconds on CPU.

For the \textit{ExploitDB Retriever}, the average processing times are 1.254 seconds on GPU and 2.377 seconds on CPU. The \textit{Metasploit Retriever} shows average times of 0.995 seconds on GPU and 1.367 seconds on CPU when processing a 40MB dense retriever. In addition, \textit{Question Retrieval System} and \textit{Entity Retriever} are noteworthy as they have the largest dense retrievers within the Structured RAG components at 3.863 GB and 554 MB respectively. The \textit{Question Retrieval System} has an average processing time of 2.536 seconds on the CPU and 2.492 seconds on the GPU. \textit{The Entity Retriever} achieves a moderate processing time of 0.250 seconds on the GPU and 0.268 seconds on the CPU due to its relatively high density.

\paragraph{\textbf{Analysis of the Performances of Unstructured RAG Components}}
As shown in Table~\ref{tab:perfomances}, GPU processing significantly increases performance, especially for components that work with large data sets. The \textsc{Text Buffers} saw significant performance gains when transitioning to the GPU, likely due to their large data size of 185 MB, which is reflected in a reduction in average time from 85.22 seconds on the CPU to 1.789 seconds on the GPU. The \textsc{Code Buffer} also benefited from GPU acceleration, but to a lesser extent, demonstrating that its tasks is less demanding on processing power. For \textsc{Metasploit Buffer}, the GPU significantly improved efficiency by managing the 186 MB dataset more effectively and reducing the average time. \textsc{Metasploit Buffer} size is different the Structured \textit{Metasploit Retriever Size} because the first contains the entire codes. The \textsc{Context Transformation} with the highest computational requirements showed the GPU's advantage in processing large amounts of data, a critical aspect of tasks that require fast data analysis and processing, highlighting the adaptability of the MoRSE system to hardware improvements.

Unstructured RAG components show a significant performance increase when running on GPUs, so they are preferred to run on GPU architectures, but only when the structured RAG fails. This is to ensure that the considerable memory resources required for the large buffers of unstructured RAG are utilized and sufficient memory is retained for the generation task.

\paragraph{\textbf{Structured Retrievers Failure Rate Analysis}}
The aim is to estimate the probability of the \textit{Structured Retrievers} not finding relevant text data in response to a user query, so that a switch to the Unstructured RAG is necessary to continue the search for relevant information. To this end, we have focused on looking only at the \textit{Contextual Information} part of the final \textit{Context} (see Figure~\ref{fig:firstlayer}). We have individually multiplied the retrievers' failure rates listed in Table ~\ref{tab:perfomances}, determining a collective probability of failure at approximately 0.2569\%, equal to an empirical rate (\(\hat{p}\)) of 0.0026. When assessing reliability, we used a 95\% confidence interval~\cite{moore2019introduction}, the empirical rate, a z-score of 1.96, and the Failure Rates (Fail.Rate) from Table ~\ref{tab:perfomances}. This analysis shows that the maximum failure rate under the tested conditions does not exceed 0.46\%, which confirms the robustness of the system when processing text-based queries.

 \begin{table*}[!ht]
\centering
\caption{Performance summary of Structured and Unstructured RAG components tested on CPU and GPU (NVIDIA A100 80GB). \textit{Fail. Qs.} states for Failed Test Queries, while \textit{Tot. Qs.} states for Total Test Queries}
\label{tab:perfomances}
\begin{tabular}{@{}lccccccccccc@{}}
\toprule           
\textbf{Component} & \multicolumn{2}{c}{\textbf{Mean Time (s)}} & \multicolumn{2}{c}{\textbf{Time Std.}} & \textbf{Fail. Qs} & \textbf{Tot. Qs} & \textbf{Fail. Rate} & \textbf{Size} & \textbf{Embed.} & \textbf{Thres.} & \textbf{No. Doc.} \\
\cmidrule(lr){2-3} \cmidrule(lr){4-5}
& \textbf{GPU} & \textbf{CPU} & \textbf{GPU} & \textbf{CPU} & & & & & \\
\midrule
\multicolumn{10}{c}{\textbf{Structured RAG Retrievers}} \\
\midrule
\textit{Malware Retriever} & 0.061 & 0.110 & 0.009 & 0.013 & 6 & 100 & 6\% & 3.9MB & $\alpha$ & 0.7 & $\sim1000$\\
\textit{Metasploit Retriever} & 0.995 & 1.367 & 0.592 & 1.307 & 5 & 100 & 5\% & 40MB & $\alpha$ & 0.75 & $\sim4900$\\
\textit{ExploitDB Retriever} & 1.254 & 2.377 & 0.998 & 1.198 & 8 & 100 & 8\% & - & - & - & $\sim13000$\\
\textit{CWE Retriever} & 0.083 & 0.108 & 0.056 & 0.019 & 28 & 100 & 28\% & 5.86MB & $\alpha$ & 0.73 & $\sim1000$\\
\textit{MITRE Retriever} & 0.057 & 0.13 & 0.07 & 0.029 & 19 & 100 & 19\% & 3.2MB & $\alpha$ & 0.7 & $\sim800$\\
\textit{Entity Retriever} & 0.250 & 0.268 & 0.078 & 0.018 & 23 & 100 & 23\% & 554MB & $\beta$ & 0.5 & $\sim7000$\\
\textit{Question Ret. Sys.} & 2.492 & 2.536  & 0.298 & 0.338 & 21 & 100 & 21\% & 3.863GB & $\beta$ & 0.6 & $\sim7000$\\
\midrule
\multicolumn{10}{c}{\textbf{Unstructured RAG Components}} \\
\midrule
\textsc{Paper Buffers} & 0.838 & 39.8 & 0.139 & 2.94 & 1 & 100 & 1\% & 86.4MB & $\beta$ & Top 5 & $\sim380$\\
\textsc{Text Buffers} & 1.789 & 85.22 & 0.298 & 6.29 & 1 & 100 & 1\% & 185MB & $\alpha$ & Top 5 & $\sim6000$\\
\textsc{Code Buffers} & 0.416 & 19.81 & 0.069 & 1.46 & 8 & 100 & 8\% & 43.0MB & $\alpha$ & Top 5 & $\sim1000$\\
\textsc{Metasploit Buffers} & 1.799 & 85.68 & 0.299 & 6.32 & 5 & 100 & 5\% & 186MB & $\alpha$ & Top 5 & $\sim4900$\\
\textsc{Context Transf.} & 4.84 & 230.5 & 0.805 & 17.01 & - & 100 & - & - & $\alpha$ & 0.6 & -\\
\bottomrule
\end{tabular}
\end{table*}

\section{Related Work}\label{sec:rw}
We now overview recent developments in Named Entity Recognition (NER), Knowledge Graphs (KGs), and Large Language Models (LLMs) that have contributed to more sophisticated, automated, and adaptive cybersecurity systems. 

\paragraph{\textbf{Named Entity Recognition in Cybersecurity}} Significant progress has been made in the evolving landscape of Named Entity Recognition (NER) for cybersecurity. In particular, the use of BERT and its Whole Word Masking variant with a BiLSTM-CRF framework has shown remarkable improvements in entity recognition metrics~\cite{zhou2021named}. Similarly, by fusing rule-based, dictionary-based methods and CRF, the RDF-CRF model has significantly improved entity recognition in the cybersecurity domain~\cite{yi2020cybersecurity}. In addition, a hybrid model that combines deep learning with dictionary-based methods has significantly improved precision and recognition in the identification of complex entities~\cite{gao2021data}. A study by Srivastava et al.~\cite{srivastava2023study} highlighted the differential effectiveness of word embeddings such as fastText, GloVe and BERT, with fine-tuned BERT embeddings with a feed forward network achieving an F1 score of 0.974, highlighting the importance of model adaptation to specific domains. In addition, the introduction of the JCLB model, which combines contrastive learning with a Belief Rule Base~\cite{hu2024joint}, showed improved accuracy through semantic expansion and optimized BRB parameters. Li et al. developed NEDetector~\cite{li2021nedetector}, which improves NER by identifying cybersecurity neologisms with 89.11\% accuracy, outperforming traditional trending tools in detecting threats on platforms like Twitter. Extractor distills attack behavior from CTI reports into clear, actionable insights and leverages provenance graphs to improve cyber analytics in threat hunting with real-world effectiveness~\cite{satvat2021extractor}. Koloveas et al.~\cite{koloveas2021intime} created the inTIME framework, which leverages machine learning to transform web data into actionable CTI, streamlining the intelligence lifecycle with a unified platform for intelligence collection, analysis and sharing. At the same time, a new threat modeling language has been developed by Xiong et al.~\cite{xiong2022cyber} based on the MITRE Enterprise ATT\&CK Matrix that integrates key elements of enterprise security to improve defense strategies through simulations. Husari et al.~\cite{husari2017ttpdrill} further refines CTI analysis by automating the extraction of threat actions from unstructured texts, employing NLP and IR for semantic extraction and aligning attack patterns with standards like STIX 2.1, achieving significant precision and recall in its evaluations.

\par MoRSE goes beyond current NER technologies by providing dynamic entity recognition and response generation. It identifies named entities in user queries in real time and uses this information to generate precise, contextualised responses. 

\paragraph{\textbf{Knowledge Graphs for Cybersecurity}}
Knowledge graphs (KGs) are revolutionizing cybersecurity, from threat intelligence to education. Agrawal et al.~\cite{agrawal2022building} have shown how KGs from unstructured text enhance cybersecurity learning, with student feedback highlighting improved understanding and engagement. Sewak et al. developed CRUSH~\cite{sewak2023crush}, which integrates Large Language Models (LLMs) such as GPT-3.5/GPT-4/ChatGPT with Enterprise Knowledge Graphs (EKGs) to create Threat Intelligence Graphs (TIG) that achieve up to 99\% recall in identifying malicious scripts.
Li et al. developed AttacKG~\cite{li2022attackg}, which automates the extraction of attack techniques from CTI reports into structured knowledge graphs, which greatly improves the analysis of attack patterns with high accuracy and supports advanced threat detection efforts.
Liu et al.~\cite{liu2022tricti} use NLP to convert over 29k cybersecurity reports into 113,543 actionable Cyber Threat Intelligence (CTI) points by highlighting campaign triggers for better classification accuracy. Piplay et al.~\cite{piplai2020creating} describes a system for generating Cybersecurity Knowledge Graphs (CKGs) from After Action Reports (AARs) using a `Malware Entity Extractor' and neural networks that improves security analysis through refined query responses. Gao et al.~\cite{gao2020hincti} use of a heterogeneous information network (HIN) and meta-path approach within a graph convolutional network is characterized by its sophisticated threat type identification capabilities validated by real-world data. Sikos et al.~\cite{sikos2023cybersecurity} discuss how knowledge graphs help in cybersecurity threat intelligence and automated reasoning, and highlights their importance in analyzing cyber data. Ren et al.~\cite{ren2022cskg4apt} present a knowledge graph for APT attack mapping that combines deep learning with network defense expertise. Mitra et al.~\cite{mitra2021combating} augment CKGs with provenance information to combat fake cybersecurity information and ensure data reliability.

\par In comparison, MoRSE addresses the limitations of Knowledge Graphs (KGs) in cybersecurity by providing dynamic updates with real-time adjustments as opposed to the manual revisions of KGs. It also provides interactive query capabilities that are real-time and interactive as opposed to the static KGs. In addition, MoRSE offers improved customizability and modularity, which allows for greater adaptability and easier expansion of the knowledge base compared to the fixed structures of KGs.

\paragraph{\textbf{LLMs and Chatbots in Cybersecurity Landscape}}
Research into chatbots in cybersecurity highlights their role in education, ethics and regulation. Yoo et al. (2024) and Abu-Amara et al. (2024) explore the impact of GDPR and the use of gamified chatbots in education, while Pieterse (2024) evaluates the utility of ChatGPT in guiding students through cybersecurity CTF challenges, noting that it reaches its limits in providing direct solutions~\cite{yoo2024lessons, abu2024spreading, pieterse2024friend}. Mitra et al.~\cite{mitra2024localintel} developed LOCALINTEL, which presents an automated system that uses LLMs to create organisation-specific threat intelligence from global and local databases to improve the efficiency of SoC operations. Juttner et al.~\cite{juttner2023chatids} use ChatGPT to make IDS alerts understandable to non-experts, improving network security in the home and home office environment. However, concerns about trust, privacy and ethics highlight the need for further research before widespread adoption. Yoo et al.~\cite{yoo2022icsa} use CNN classifiers and AI chatbot to detect and combat SNS phishing attacks before they can occur. This is more promising than traditional methods because it provides real-time support and actions on Telegram, with validated effectiveness against LSTM models.
Iqbal et al.~\cite{Iqbal2023} explore the dual utility of ChatGPT in cybersecurity, highlighting its benefits for defense strategies and the risks of its misuse in cyberattacks, and call for more research on its offensive capabilities. Chamberlain and Casey~\cite{chamberlain2024capture} examine the application of ChatGPT in penetration testing and CTF exercises, pointing to its potential to create dynamic scenarios and enhance the learning process. Aghaei et al.~\cite{aghaei2022securebert} developed SecureBERT, which automates critical cybersecurity tasks by introducing a specialized language model for Cyber Threat Intelligence (CTI) that uses a customized tokenizer and pre-trained weights for improved performance on NLP tasks. Ameri et al.~\cite{ameri2021cybert} use BERT for feature classification and achieves improved accuracy from 76\% to 94.4\% by optimizing the hyperparameters. It demonstrated robustness with a standard deviation of ±0.6\% across all validations and outperformed models such as GPT2, ULMFiT, ELMo, CNN, LSTM and BiLSTM on cybersecurity tasks. Voros et al.~\cite{voros2023web} use knowledge distillation from LLMs to efficiently categorize URLs, reduce the number of parameters and improve inline scanning. Happe et al.~\cite{happe2023getting} use GPT-3.5 to extend penetration testing and demonstrates LLMs as AI sparring partners in security testing. Lu et al.~\cite{lu2024grace} integrate graph structural information and in-context learning into LLM-based software vulnerability detection, significantly outperforming conventional models. Yu et al.~\cite{yu2023honey} use GPT-3 to generate semantic honeywords containing users' PII, which improves their indistinguishability and increases defenses against security breaches.

\par Unlike traditional LLMs and chatbots, MoRSE stands out by providing instant access to a constantly updating cybersecurity knowledge base, quickly integrating the latest threats and solutions, and enabling extensive customization for various cybersecurity needs. MoRSE does not focus on niche cybersecurity topics; instead, it aims to provide comprehensive coverage of cybersecurity knowledge. In addition, MoRSE enhances the user experience by providing user-friendly access to complex cybersecurity information, increasing flexibility and expanding accessibility.

\section{Conclusions and Future Work}\label{sec:cf}
With cyber threats on the rise, effective cybersecurity strategies are becoming increasingly important. Integrating continuous learning and Retrieval Augmented Generation (RAG) into large language models (LLMs) improves their accuracy and timeliness in responding to these threats.
In this paper, we have investigated the use of two Retrieval Augmented Generation (RAG) systems, namely \textit{Structured} RAG and \textit{Unstructured} RAG, to provide precise and structured answers to cybersecurity queries. In \textit{Structured RAG}, we have focused on implementing parallel retrievers to quickly and effectively find the relevant document in response to a user query. \textit{Unstructured RAG}, on the other hand, is designed to answer even the most complicated cybersecurity queries. We have implemented an evaluation suite to assess the relevance, similarity, and correctness of the answers generated by our system compared to ground truth. The performance of our system was compared to other renowned commercial Large Language Models (LLMs) using two additional test suites that adhere to the \textit{LLM as a Judge} paradigm. The results show that our system outperforms  competing models by more than 10\% in terms of the correctness and relevance of the given answers and by 50\% in terms of accuracy against GPT-4 for questions related to vulnerabilities.

In order to make the framework publicly available, our goal is to improve MoRSE with a Prefix-aware Greedy replacement policy (PAGRP)~\cite{jin2024ragcache} for the semantic cache. This takes into account the initial segments of queries or data to make more informed decisions about what data to store in the cache. The PAGRP prioritizes the cache items based on their access frequency and size, ensuring that the system stores the most useful data and minimizes the likelihood of cache misses. In addition, we plan to replace the MITRE retriever with a comprehensive Knowledge Graph. This Knowledge Graph will include mitigation and detection approaches and aggregate real malware reports associated with the corresponding MITRE software. This solution enables better threat insights by allowing the computation of community analysis, centrality algorithms, and threat similarities quickly and in real-time.


%
\bibliographystyle{IEEEtran}
\bibliography{bibliography}

\end{document}